\documentclass[%
 reprint,
 amsmath,amssymb,
 aps,
]{revtex4-2}

\usepackage{graphicx}
\usepackage{dcolumn}
\usepackage{bm}
\usepackage{mathrsfs,amsmath}

\begin{document}

\preprint{APS/123-QED}

\title{Broadband time-modulated absorber beyond the Bode-Fano limit by energy trapping}

\author{Xiaozhen Yang}
\email{xiy003@eng.ucsd.edu}
\author{Erda Wen}%
\email{ewen@eng.ucsd.edu}
\author{Daniel F. Sievenpiper}%
\email{dsievenpiper@eng.ucsd.edu}
\affiliation{%
 Applied Electromagnetics Group, Electrical and Computer Engineering Department,\\
 University of California, San Diego, California 92093, USA
}%


\date{\today}

\begin{abstract}
Wide-band absorption is a popular topic in microwave engineering to protect sensitive devices against broadband sources. However, the Bode-Fano criterion defines the trade-off between bandwidth and efficiency for all passive, linear, time-invariant systems. In this letter, we propose a broadband absorber beyond the Bode-Fano limit by creating an energy trap using time-modulated switch/diodes. This work starts with an ideal circuit model to prove the concept, followed by two EM realizations - a freuqnecy selective surface (FSS) approach for general bandwidth broadening and a low-profile PCB design. The prototype of the latter is built and measured, demonstrating a Bode-Fano integral larger than one. This approach paves a way to many practical ultra-wide band absorber designs. 
\end{abstract}

\maketitle

\begin{figure*}
\includegraphics{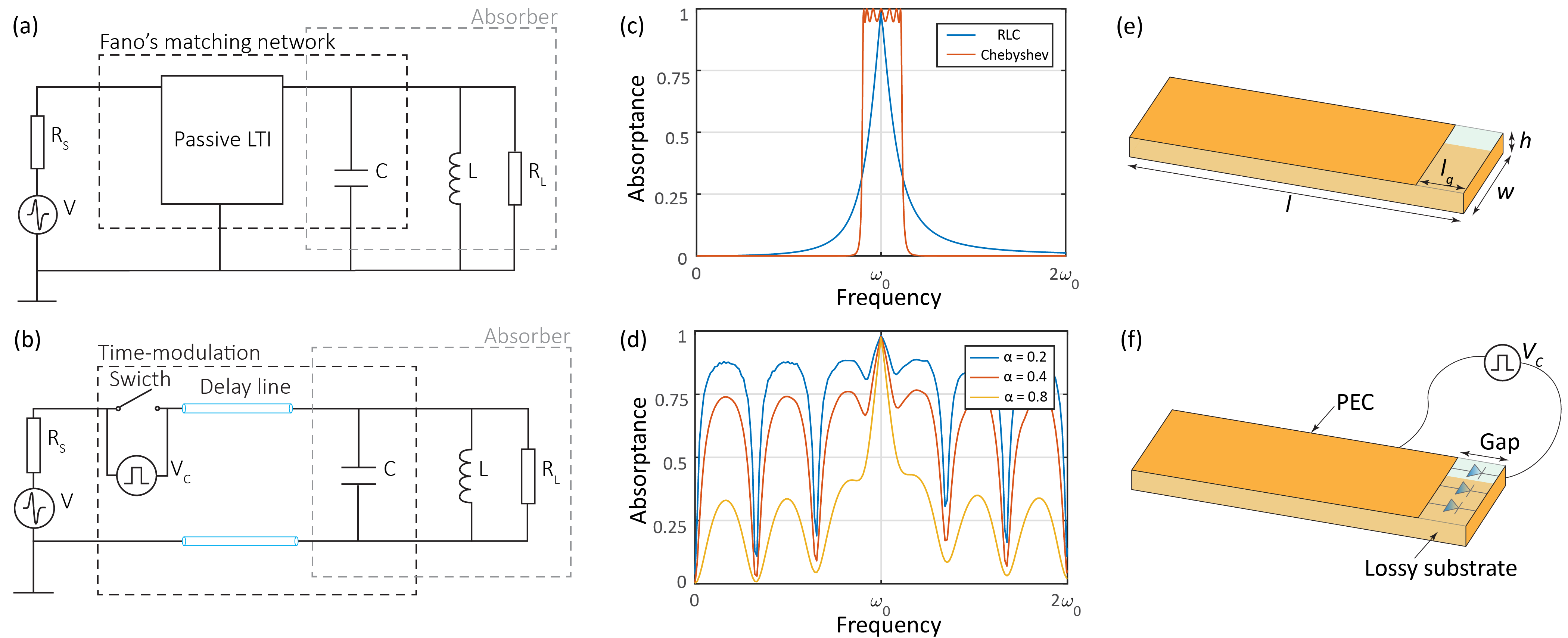}
\caption{\label{fig:concept}(a) Circuit representing a conventional single-resonance absorber, with $R_S=R_L=653\;$Ohms, $C=0.5\;$pF, $L=3\;$nH. (b) Circuit of the proposed time-modulated absorber. (c) Absorptance spectrum of (a) without matching and with a $7^{th}$ order Chebyshev bandpass filter, respectively, $\omega_0=1/\sqrt{LC}$. (d) Absorptance spectrum of (b), with $t_d=0.37\;$ns through $15$ reflections. (e) EM model of a conventional absorber being measured, with $h=0.5\;$cm, $w=2.5\;$cm, $l=7\;$cm, $l_g=1\;$cm. The substrate is FR4 with $\epsilon_r=4.8$. (f) EM model of the proposed time-modulated absorber being measured.}
\end{figure*}

\textit{Introduction.---}Wide-band absorption has aroused attention in microwave engineering \cite{Pozar11_FUND,Collin07_FUND,Balanis99_FUND,Balanis15_FUND,Sedra98_FUND} during recent years to avoid damage from high energy and broadband microwave sources \cite{Kopp96_EB1,Kopp96_EB2}. Related research from metamaterials to nanostructures \cite{Cui11_ABS,Zhu20_ABS} are conducted in attempt to provide absorption covering the range from microwave frequencies \cite{Yang07_ABS,Zhang19_ABS} to infrared \cite{Lei18_ABS,Liu20_ABS} and Terahertz \cite{Zhu14_ABS,Biabanifard18_ABS}. Conventional methods focus on optimizing the balance between structure size and bandwidth by adopting passive multi-resonance structures \cite{Chen13_MR,Cui11_ABS}. However, there is a physical limit between absorptance and bandwidth for any passive linear time-invariant(LTI) system, the Bode-Fano limit \cite{Bode45_BF,Fano50_BF,Kerr95_BF}, which describes the trade-off between matching bandwidth and efficiency and is further proved to be applicable to antennas \cite{Ghorbani06_BF1,Ghorbani06_BF2}, in another form, the Rozanov bound for absorbers \cite{Rozanov00_BANDWIDTH}. Fig.\ref{fig:concept}(a) illustrates an example of a single-resonance absorber represented by a parallel $RLC$ circuit with a broadband $3^{rd}$ order derivative Gaussian pulse input covering from DC to around $8\;$GHz, whose center frequency aligns with the resonating frequency of the $RLC$ circuit. The absorptance on the load $R_L$ is shown in Fig.\ref{fig:concept}(c) when the matching network is either void or a $7^{th}$ order Chebyshev bandpass filter (See supplementary I). It can be observed that while absorptance approaches 1 around the center frequency when the Chebyshev filter is applied, the bandwidth is also compromised, and as expected, the integral under the both two curves fall below the Bode-Fano limit,
\begin{equation}
\frac{R_{L}}{\pi L}\int_{0}^{\infty} \frac{1}{\omega^2}\ln{\frac{1}{|\rho|}}d\omega \leq 1\label{eq:BF for LTI}
\end{equation}
where $\rho$ is the reflection coefficient.

One possible solution to achieve beyond the Bode-Fano limit is to break its constraints of being a passive LTI system by introducing active devices, nonlinearity, or time-dependent elements, such as switches and diodes, etc. \cite{Shlivinski18_BF}. Non-foster components can break the passivity constraint by integrating amplifiers and other active elements, achieving ultra-wide band matching in contrast with conventional matching networks \cite{Sussman09_NF,Chen13_NF,Shi19_NF}. Additionally, research in direct antenna modulation (DAM) indicates that by applying time-modulation, transmitters can radiate beyond their original bandwidth \cite{Jing14_DAM,Yoa04_DAM1,Yao03_DAM2}. Recent research on time-modulated absorbers \cite{Firestein21_Rozanov,Li21_Bandwidth} has developed analysis of absorbers with time-varying electromagtic properties, such as the conductivity, permittivity, and permeability, and discuss a modulation strategy to go beyond the Rozanov bound. However, it is difficult to realize an absorber with varying dielectric characteristics in the real world, which hinders their applications.

In this letter, we propose a practical approach to break the Bode-Fano limit of conventional absorbers by applying temporal modulation to integrated diodes/switches based on an energy trapping concept. The idea is to trap the incident ultra-wide band pulses within the resonator by adopting a modulated switch that cuts off the radiation path right after the signal fully enters the structure. As a proof of concept illustrated in Fig.\ref{fig:concept}(b), the switch, along with a delay line with a delay time of $t_{d}=t_{p}/2$, are inserted between the same Gaussian source and absorber as in Fig.(\ref{fig:concept})(a). The switch has a turn-on time of $t_{p}$, synchronized with the pulse, and is then left off for the rest of the period, so that the input energy is allowed to only enter, but not to leave the network. As a result, the energy dissipates inside the network through multiple reflections. To obtain the transient response on $R_{L}$, each reflection is treated separately using S parameters and summed up with its corresponding time shift (for details see Supplementary II).
\begin{equation}
R_{tot}(t)=\sum\limits_{n=1}^N R_{n}(t-2n t_{d})
\label{eq:AnalyticalTimeShift}
\end{equation}
\begin{equation}
R_{n}(t)=\mathscr{F}^{-1}\left\{ S_{21}(f) \mathscr{F}\left\{p_{n}(t)\right\}\right\}
\label{eq:AnalyticalResponse}
\end{equation}
\begin{equation}
p_{n}(t)=\mathscr{F}^{-1}\left\{(S_{11})^{n-1}\mathscr{F}\{p(t)\}(1-\alpha)^{n-1})\right\}
\label{eq:AnalyticalInput}
\end{equation}
where $R_{n}$ represents the transient response on the load during the $n^{th}$ reflection neglecting the time delay, $p_{n}$ is the input of the $n^{th}$ reflection, $S_{ii}$ is the S parameters of parallel $LC$ network, and $p(t)$ is the original input from the broadband source. Here we define the imperfection of the switch by a leakage factor $\alpha=E_{Trans.}/E_{Inc.}$. If $\alpha=0$, all waves are trapped in the $RLC$ circuit. Otherwise part of the incoming energy leaks out due to the imperfection of the switch.  Fig.\ref{fig:concept}(d) shows the absorptance for a time-modulated absorber circuit model with different $\alpha$. Note that the absorptance across the whole spectrum is 1, assuming an ideal switch with $\alpha=0$ is used. Even with a highly imperfect switch, a drastic increase of bandwidth is observed in Fig.\ref{fig:concept}(d) compared with (c), breaking the Bode-Fano limits even when the upper limit of the Bode-Fano integral is chosen to be a finite value such as $2\omega_{0}$: as an example, for $\alpha=0.4$, ${R_{L}}/{\pi L}\int_{0}^{2 \omega_{0}} \frac{1}{\omega^2}\ln{\frac{1}{|\rho|}}d\omega=30.4$.

The most straightforward way to realize the above circuit in an electromagetic structure is by adopting diode-integrated frequency selective surfaces (FSS) as the switch, and free space as the delay line (see Supplementary III). Thus, the ON and OFF states of the FSS corresponds to the ON and OFF states of the diodes. With the FSS in the OFF state, incident waves are allowed to propagate through; while in the ON state, the energy is trapped between the FSS and the conventional absorber. This method is rather general and can be applied to any passive absorber to broaden its bandwidth. However, the increased space needed conflicts with the objective of broadband absorption given a limited thickness. Thus, in this letter, a more practical low-profile PCB realization is shown in Fig.\ref{fig:concept}(e) and (f). A horizontally propagating incident pulse/pulse train is coupled to the structure through the gap on the top surface, the length of the structure performs equivalently as the delay line in the circuit model, and the switch is replaced by step recovery diodes (SRD) contolled by a signal $V_{C}$. Instead of a lumped resistor $R_{L}$, the dissipation is distributive in the lossy FR4 substrate. When the control signal $V_C$ is low, the diodes are OFF, and energy couples into the absorber; when $V_C$ is high, the diodes turn ON, the absorber behaves as a closed cavity, and all the energy is trapped inside. The simulated static ON and OFF states results are included in Supplementary IV.

\begin{figure}
\includegraphics{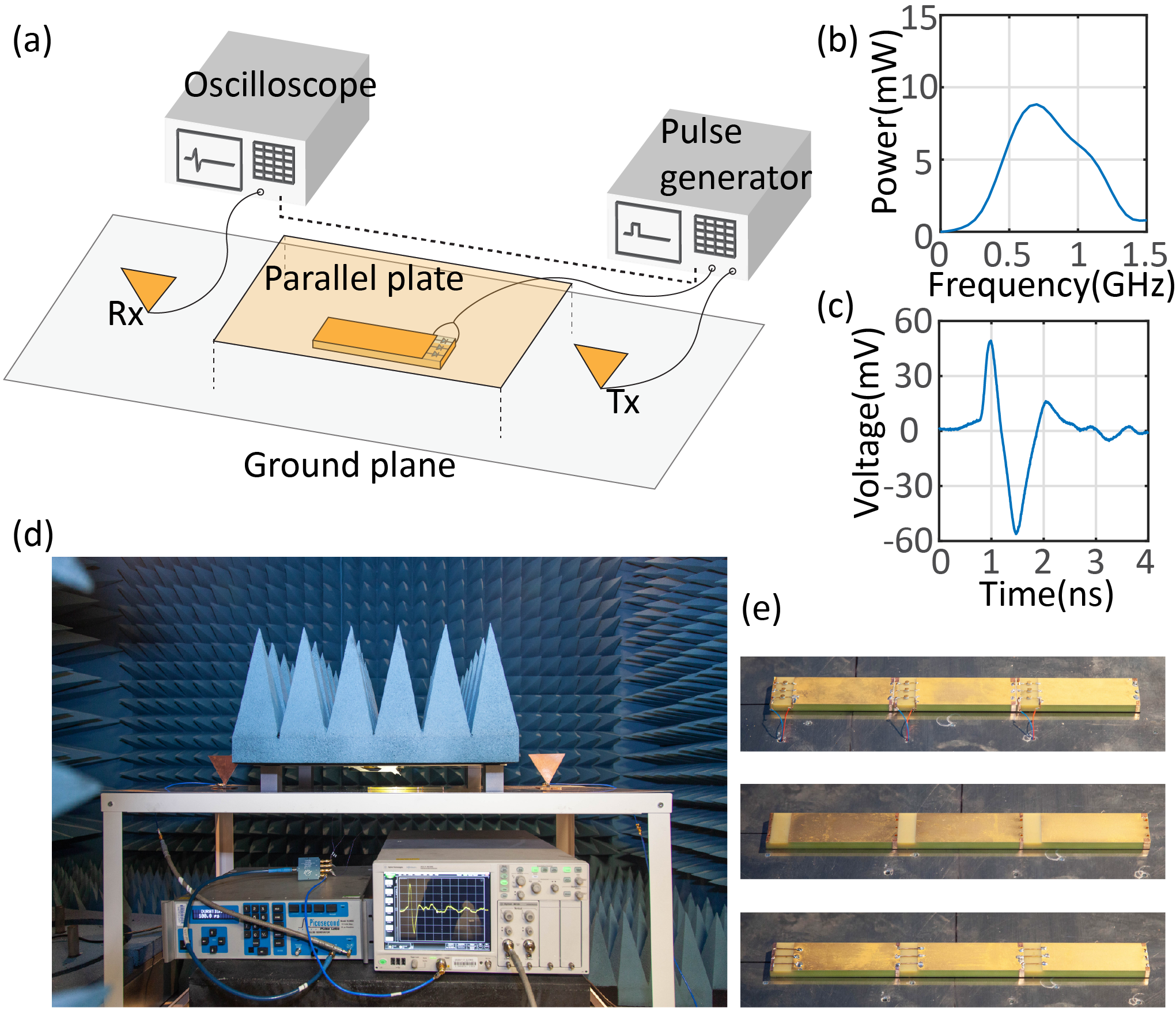}
\caption{\label{fig:measurement}(a) Measurement setup.(b) Measured spectrum of input signal from pulse generator.(c) Measured transient waveform of input signal from pulse generator. (d) Picture of measurement setup in the anechoic chamber. (e) Measurement time-modulated, OFF state and ON state samples.}
\end{figure}

\begin{figure*}
\includegraphics{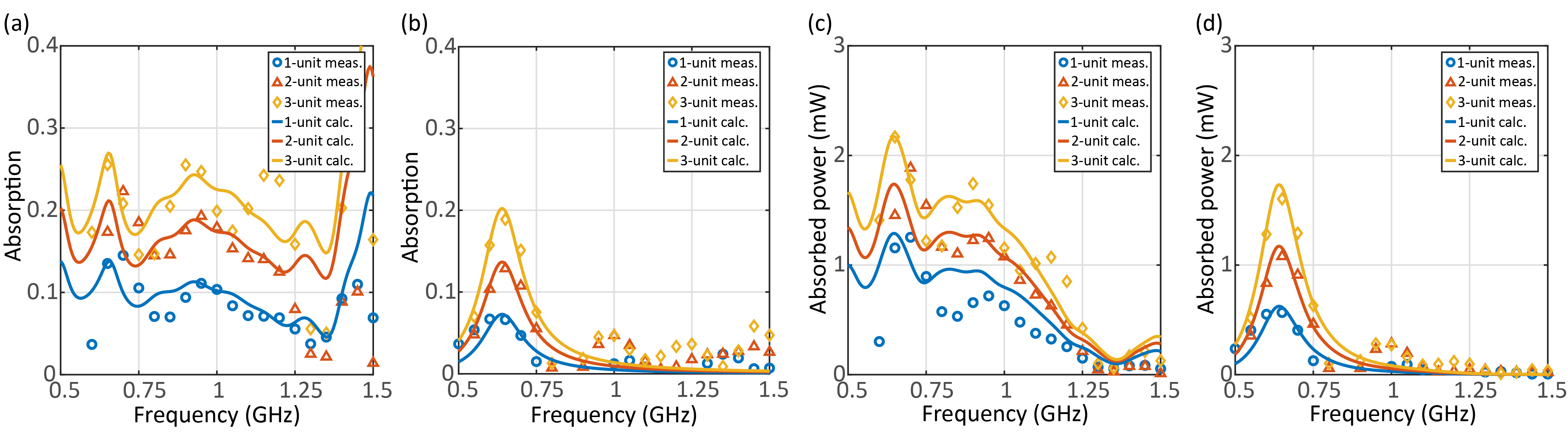}
\caption{\label{fig:results}(a) Measured and calculated absorptance for time-modulated one-, two- and three-unit sample, respectively. $r$ is assumed to be $0.27,\;0.37,\;0.47$ in calculation for 1-,2-and 3-unit, respectively. $\alpha=0.3$ for all calculations. (b) Measured and calculated absorptance for non-modulated one-, two- and three-unit sample, respectively. (c) Measured and calculated absorbed power for time-modulated one-, two- and three-unit sample, respectively. (d) Measured and calculated absorbed power for non-modulated one-, two- and three-unit sample, respectively. }
\end{figure*}

\textit{Measurement.---} The measurement sample is shown in Fig.\ref{fig:measurement}. The size of the structure is the same as shown in Fig.\ref{fig:concept}(e) and (f), and the measurement setup is illustrated in Fig.\ref{fig:measurement}(a). The non-modulated sample with static ON and OFF states, and the time-modulated sample are measured, respectively, for one, two and three units. The signal is generated by a fast rising pulse generator and radiated by a broadband monopole bow-tie antenna Tx sitting on a $120\;$cm by $60\;$cm ground plane, travelling through a parallel plate $60\;$cm by $40\;$cm by $6\;$cm TEM waveguide (to confine energy) with the sample-under-test inside. The transmitted signal is received by another bow-tie antenna Rx and collected by a wide-band oscilloscope. The produced signal from the pulse generator is a $5\;$V repetitive pulse train lasting $100\;$ps with low duty cycle of $0.001\%$ and short rise time around $70\;$ps. This signal is modulated by Tx forming a $3^{rd}$ order derivative Gaussian-like pulse, collected by Rx. The spectrum and transient waveform is shown in Fig.\ref{fig:measurement}(b) and (c). The distance between Tx and Rx is $90\;$cm. The osilloscope and the pulse generator are synchronized with the trigger signal of the latter, and part of the trigger signal is split out by a power divider to be used as the control signal for the time-modulated absorber. The trigger signal has an amplitude of around $5\;$V and lasts for $40\;$ns, which warrants adequate multi-reflection for sufficient dissipation. Short twisted wires with different lengths are adopted and tuned to precisely control the time delay between different units for multi-unit measurements when travelling wave modulation is performed. All the cables involved in this experiment are carefully chosen, such that the modulated signal synchronizes with the incident wave for the samples to trap as much energy as possible. For each configuration, the average of five sets of measurements is adopted to reduce noise and avoid fluctuations. 

From an analytical perspective, the $R$, $L$, $C$ and $R_{S}$ required for S parameter calculation in (\ref{eq:AnalyticalInput}) can be extracted from non-modulated measurement results in Fig.\ref{fig:results}(b) by
\begin{equation}
L=\frac{1}{2\pi}\frac{\Delta f}{f_{0}^2}R
\label{eq:L}
\end{equation}
\begin{equation}
C=\frac{1}{2\pi R\Delta f}
\label{eq:C}
\end{equation}
where $f_{0}=640\;$MHz is the center frequency, $\Delta f=80\;$MHz is the $3\;$dB bandwidth. The resistance is chosen to be the characteristic impedance of free space $R=R_{L}=R_{S}=377\;$Ohms.

Different from the circuit model in Fig.\ref{fig:concept}(b), only a portion of radiation from Tx couples into the sample in the measurement. Therefore, a parameter $r$ is introduced to describe how much energy is captured by the units and Eq.(\ref{eq:AnalyticalInput}) accordingly changes to
\begin{equation}
p_{n}(t)=r\mathscr{F}^{-1}\left\{(S_{11})^{n-1}\mathscr{F}\{p(t)\}(1-\alpha)^{n-1})\right\}
\label{eq:ModifiedInput}
\end{equation}
Moreover, since the loss is distributive, instead of evaluating the absorption by calculating the dissipation on $R_L$, the absorptance $A$ and absorbed power $P$ is calculated with respect to transmission $T$ instead of absorption $R_{tot}$ on the load by
\begin{equation}
A_{cal.}(f)=\frac{\mathscr{F}\left\{p_{t}\right\}^2-\mathscr{F}\left\{T_{tot}(t)\right\}^2}{\mathscr{F}\left\{p_{t}\right\}^2}
\label{eq:Absorption}
\end{equation}
\begin{equation}
P_{cal.}(f)=\mathscr{F}\left\{p_{t}\right\}^2-\mathscr{F}\left\{T_{tot}(t)\right\}^2
\label{eq:AbsorbedPower}
\end{equation}
in which,
\begin{equation}
T_{tot}(t)=\sum\limits_{n=1}^N T_{n}(t-2nt_{d})
\label{eq:AnalyticalTransmissionTimeShift}
\end{equation}
corresponds to the transient transmission received by Rx and collected by the oscilloscope, and
\begin{equation}
T_{n}(t)=\alpha \mathscr{F}^{-1}\left\{\mathscr{F}\{S_{11}p_{n}(t)\}\right\}
\label{eq:AnalyticalTransmission}
\end{equation}
For the measurement results, to exclude the disturbance of fields by the samples, the baseline is the ON state (replacing diodes with wires) instead of an empty waveguide.
\begin{equation}
A_{meas.}(f)=\frac{\mathscr{F}\{T_{ON}(t)\}^2-\mathscr{F}\{T_{TM/OFF}(t)\}^2}{\mathscr{F}\{T_{ON}(t)\}^2}
\label{eq:MeasAbsorption}
\end{equation}
\begin{equation}
P_{meas.}(f)=\mathscr{F}\{T_{ON}(t)\}^2-\mathscr{F}\{T_{TM/OFF}(t)\}^2
\label{eq:MeasAbsorptionPower}
\end{equation}
where $T_{ON/OFF/TM}$ denotes transmitted signal measured with ON state, OFF state and time-modulated sample collected by the oscilloscope.

The measured and calculated absorptance is shown in Fig.\ref{fig:results}. As shown in Fig.\ref{fig:results}(b), the non-modulated sample absorbs energy at around the resonating frequency $0.64\;$GHz, while the modulated sample has a broadband absorption starting from $0.6\;$GHz to $1.5\;$GHz. Due to difficulty in directly determining $\alpha$ and $r$ through measurement, these two parameters are numerically fitted. It is found that when $\alpha=0.3$ (indicating $9\;\%$ of the coupled energy leaks out when the diodes are ON) and $r=0.27,\;0.37,\;0.47$ for one-,two- and three-unit calculations, the predicted results show a good match with the measurement results in Fig.\ref{fig:results}. The calculated Bode-Fano integral ${R_{L}}/{\pi L}\int_{0.5\;GHz}^{1.5\;GHz} \frac{1}{\omega^2}\ln{\frac{1}{|\rho|}}d\omega=5.81$ and $0.95$ for modulated and non-modulated sample, respectively. The absorption increases as the number of units increases for both modulated and non-modulated cases, indicating the potential to accommodate this design to larger-scale metasurfaces. 

\textit{Conclusion.---} In this letter, we proposed an approach to break the Bode-Fano limit using time-modulation by creating an energy trap and discussed two possible EM structures based on the idea. Verified by analytical calculations, simulation results and measurement of the horizontal model, we proved the proposed method can achieve beyond the bandwidth limitations for passive LTI systems. The latter low-profile design integrates well with PCB and can be potentially applied to protect sensitive devices against broadband signals. Although measurements are done only for up to three units, the horizontal design could be extended to a metasurface. Further research may focus on self-triggering and self-delaying modules. Time-modulation also brings other interesting properties, such as non-reciprocity \cite{Correas15_NR,Li20_NR,Cardin20_NR}. The horizontal model we propose also has non-reciprocal behaviour caused by the temporal modulation and is discussed in Supplementary IV.

The authors would like to acknowledge the support by Office of Naval Research under Grant No. N00014-20-1-2710.

\nocite{*}

\bibliography{apssamp}

\end{document}